\documentclass{article}
\usepackage{spconf,amsmath,graphicx}

\usepackage{bm}
\usepackage{multirow}
\usepackage{algorithm} 
\usepackage{algpseudocode} 
\usepackage{url}

\usepackage{breakurl}
\usepackage[breaklinks, hidelinks]{hyperref}
\usepackage[dvipsnames]{xcolor}

\PassOptionsToPackage{hyphens}{url}

\title{Learning ASR pathways: A sparse multilingual ASR model}
\name{Mu Yang$^{1, \dagger}$, Andros Tjandra$^{2}$, Chunxi Liu$^{2}$, David Zhang$^{2}$, Duc Le$^{2}$, Ozlem Kalinli$^{2}$ \thanks{$\dagger$ Work was done when Mu Yang was an intern at Meta.}}
\address{ \textsuperscript{1}Center for Robust Speech Systems (CRSS), University of Texas at Dallas, USA \\ \textsuperscript{2}Meta AI, USA \\ 
{\small \tt  mu.yang@utdallas.edu, androstj@fb.com} }
\begin{document}
\ninept
\maketitle
\begin{abstract}
Neural network pruning compresses automatic speech recognition (ASR) models effectively. However, in multilingual ASR, language-agnostic pruning may lead to severe performance drops on some languages because language-agnostic pruning masks may not fit all languages and discard important language-specific parameters. In this work, we present \textit{ASR pathways}, a sparse multilingual ASR model that activates language-specific sub-networks (“pathways”), such that the parameters for each language are learned explicitly. With the overlapping sub-networks, the shared parameters can also enable knowledge transfer for lower-resource languages via joint multilingual training. We propose a novel algorithm to learn ASR pathways, and evaluate the proposed method on 4 languages with a streaming RNN-T model. Our proposed ASR pathways outperform both dense models and a language-agnostically pruned model, and provide better performance on low-resource languages compared to the monolingual sparse models.
\end{abstract}
\begin{keywords}
Multilingual, speech recognition, sparse, pruning.
\end{keywords}

\section{Introduction}
\label{sec:intro}
Automatic speech recognition (ASR) technologies play an important role in customer smart devices such as smartphones, smart speakers, smart watches, virtual reality glasses, and more. Due to the limited computation resources and storage, hosting a high-performance yet dense ASR model on on-device hardware is challenging. In order to reduce on-device model size and boost run-time efficiency, recent works have attempted to compress the ASR models through knowledge distillation \cite{movsner2019improving, panchapagesan2021efficient}, model sparsification \cite{shangguan2019optimizing, yang2022omni, ding2021audio, liu2022learning}, quantization \cite{he2019streaming, sainath2020streaming, ding20224}, etc. In this study, we compress ASR models via neural network pruning (i.e. zeroing out a certain amount of dense model weights by a learned pruning mask). We specifically target at multilingual ASR scenario \cite{toshniwal2018multilingual, pratap20c_interspeech, li2021scaling, kannan19_interspeech, bai2022joint}, with the aim that sparsifying a multilingual ASR model does not significantly degrade recognition performance on certain languages.

Pruning a multilingual ASR model poses unique challenges. Given a dense multilingual model, simply doing \textbf{language-agnostic} pruning leads to large performance drops on certain languages. This is because language-agnostic pruning only learns a single language-agnostic mask and shares parameters for all languages, which may not fit for certain languages \cite{lu2022language, gaur2021mixture}. As a result, some languages will see significant word error rate (WER) degradation. Alternatively, one can do \textbf{language-specific} pruning on monolingual data \cite{shangguan2019optimizing, yang2022omni}, which learns different model parameters and different pruning masks across languages. However, the resulting models lose their multilingual ability: we end up with one sparse monolingual model for each language, which complicates on-device deployment.

\begin{figure}
    \centering
    \includegraphics[width=1.0\columnwidth]{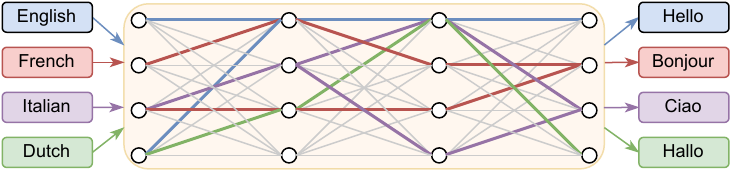}
    \caption{Demonstration of ASR pathways: each language activates its own sparse sub-network.}
    \label{fig:intro}
    \vspace{-4mm}
\end{figure}

To tackle the challenges above, we propose that the sparse multilingual model should have language-specific masks while sharing the same set of model parameters. In other words, as shown in Figure \ref{fig:intro}, we aim to obtain a multilingual model that is sparsely activated, forming sparse ``pathways" \cite{dean_2021} (sub-networks) which can be dedicated to different languages. For this purpose, we propose a novel 2-step training scheme, where the first step identifies the language-specific masks via Iterative Magnitude Pruning (IMP) \cite{zhu2017prune} or Lottery Ticket Hypothesis (LTH) \cite{frankle2018lottery}, and in the second step, we fix the masks and update the model parameters via multilingual training, where each language governs the updates of the parameters under the mask that corresponds to the input language. Through the learning process, the model can learn the optimal pathway for each language and also learn the shared parameters across languages - such that when the language-specific masks applied, the model can recognize each language independently. 

We applied the proposed method to an efficient memory transformer (Emformer) based \cite{shi2021emformer, shi2022streaming} low-latency streaming RNN-T architecture \cite{graves2012sequence} with a 70.6\% block-wise structured sparsity \cite{narang2017block}. Our empirical results on a 4-language dataset show that the proposed multilingual ASR pathways model outperforms both dense model (-5.0\% average WER) and language-agnostic pruning (-21.4\% average WER), and has better performance on low-resource languages compared to the monolingual sparse models. By analyzing the language-specific masks, we reveal that with the help of a regularization that is tailored to structured pruning, LTH masks are superior to IMP masks, even with less total effective parameters.

\section{Related works}
\textbf{Multilingual ASR.} Among recent works on multilingual ASR, language-specific modeling is usually adopted to achieve universally decent performance on all languages. This includes language-aware encoding \cite{tian2022lae}, parameterizing language-specific attention heads \cite{zhu20d_interspeech}, adapter modules \cite{kannan19_interspeech}, decoders \cite{pratap20c_interspeech} and Mixture-of-experts \cite{gaur2021mixture}. In \cite{lu2022language}, LTH was used to identify language-specific sub-networks inside the pre-trained multilingual XLSR model \cite{conneau2020unsupervised}. These sub-networks were shown to be able to pre-train an enhanced XLSR model that improves downstream multilingual ASR fine-tuning. In contrast to \cite{lu2022language}, we directly identify the language-specific sparse structures in a multilingual ASR model, which can potentially learn masks that are more tailored to ASR tasks.

\noindent \textbf{Model compression.} Prior works on ASR model compression can be grouped into 3 categories: (1) knowledge distillation \cite{movsner2019improving, panchapagesan2021efficient} which transfers knowledge from a larger model to a smaller model; (2) model sparsification \cite{shangguan2019optimizing, yang2022omni, ding2021audio, liu2022learning, wu2021dynamic} which prunes (zeros out) a subset of model parameters to reduce the representational complexity and improve run-time efficiency; (3) parameter quantization \cite{he2019streaming, sainath2020streaming, ding20224} which represents the trained model parameters with fewer bits for model size reduction. In this study, we adopt model pruning approaches (IMP and LTH) to find language-specific sub-networks. In \cite{ding2021audio}, LTH was shown to learn a better sparse monolingual ASR than IMP. We study how the monolingual sub-networks could be used for multilingual scenario. A similar scheme was also investigated in \cite{sun2020learning} for multitask sequence-labeling.


\section{Proposed method}

\begin{figure}
    \centering
    \includegraphics[width=0.95\columnwidth]{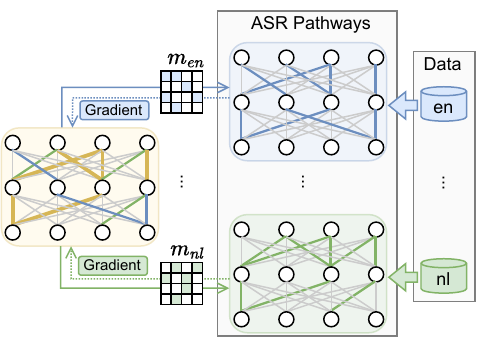}
    \caption{Multilingual training with language-specific masks. $m_{en}$ and $m_{nl}$ denotes English (\textit{en}) and Dutch (\textit{nl}) mask, respectively. Blue and green segments represent trainable weights for \textit{en} and \textit{nl}. Bold yellow weights are shared between \textit{en} and \textit{nl}. Grey weights are set to zero during training and inference.}
    \label{fig:method_pathway}
    \vspace{-3mm}
\end{figure}
Firstly, we introduce 2 pruning methods: IMP and LTH, which we use to learn language-specific masks (Section \ref{sec:imp_lth}). We then show how these identified masks are utilized to train our proposed ASR pathways (Section \ref{sec:pathway}). Finally, we describe an essential regularization technique to improve the pruning performance (Section \ref{sec:lasso}).

\subsection{Learning language-specific pruning masks} \label{sec:imp_lth}

\subsubsection{Iterative Magnitude Pruning (IMP)} \label{sec:imp_alg}
For a network $f(x; \theta)$ with input samples $x$ and model weights $\theta$, we denote a sub-network as $f(x;m \odot \theta)$, where $m \in \{0,1\}^{|\theta|}$ is a binary pruning mask, and $\odot$ represents element-wise product. In this work, we consider a structured $8 \times 1$ block-wise pruning strategy \cite{yang2022omni} which means that weight matrices will be zeroed out in $8 \times 1$ groups. We prune the encoder Emformer layers and the predictor LSTM layer, with a uniform sparsity across all prunable layers \cite{yang2022omni, liu2022learning}. Given the pre-trained dense weights $\theta_0$, we initialize $\theta = \theta_0$ and $m = 1^{|\theta_0|}$, and proceed as follows:

\noindent \textbf{Repeat}
\begin{enumerate}
    \item Train $f(x;m \odot \theta)$ for $T$ steps, leading to $f(x;m \odot \theta_T)$.
    \item Prune $p\%$ of the remaining weights that have the smallest magnitudes. Update $m$ by setting the pruned positions to 0.
    \item Set $\theta = \theta_T$ (start from the updated weights in next iteration).
\end{enumerate}
\noindent \textbf{Until} $m$ reaches the target sparsity. 

\noindent After $n$ pruning iterations, $(1-p\%)^n$ weights survive. When the target sparsity is reached, $m$ will be fixed afterward and we continue training until convergence. Note that depending on the training data, IMP can be language-agnostic (multilingual data) or language-specific (monolingual data). If trained on monolingual data, we consider the final mask $m$ as the language-specific mask. We set $p=20$, i.e. 20\% of the remaining weights will be pruned in each iteration. Pruning interval $T$ depends on training data size.


\vspace{-2mm}
\subsubsection{Lottery Ticket Hypothesis (LTH)} 
LTH aims to identify a sub-network from a random or pre-trained initialization $\theta_0$, which can be trained or transferred in isolation and obtain matching performance as the original dense network \cite{frankle2018lottery, chen2020lottery}. The algorithm for identifying such a sub-network is identical to IMP, except that there is a ``rewinding" step replacing step 3 of the pruning loop in Section \ref{sec:imp_alg}. In LTH, the rewinding step always sets $\theta = \theta_0$, as opposed to $\theta = \theta_T$ in IMP, in each iteration. After rewinding, LTH may learn a set of model parameters that are different from IMP after training on the same data for certain steps, which decide different weights to be pruned. Hence, the final mask $m$ identified by LTH may also be different from that identified by IMP. In \cite{ding2021audio}, the authors show that the sparse network identified by LTH can outperform IMP on a monolingual ASR task. In this work, we also compare the performance of the multilingual ASR pathways trained using masks identified by IMP and LTH, respectively.

\subsection{Multilingual ASR pathways training} \label{sec:pathway}
With the language-specific masks separately learned on each language, the ASR pathways are trained in a multilingual manner. Figure \ref{fig:method_pathway} demonstrates a 2-language case. Given the pre-trained dense multilingual ASR model weights $\theta_0$, and the language-specific masks $m_{en}$ and $m_{nl}$ learned via IMP or LTH, we identify the language-specific sub-networks $f(x;m_{en} \odot \theta_0)$ and $f(x;m_{nl} \odot \theta_0)$. During training, for each step, we sample a monolingual batch from the multilingual dataset, identify a language-specific sub-network using the mask corresponding to the input language, do a forward and backward pass in the sub-network, and only update the parameters inside the sub-network. After training, we end up with a set of universal weights $\theta^*$, and $f(x;m_{en} \odot \theta^*)$ and $f(x;m_{nl} \odot \theta^*)$ form the ASR pathways for English and Dutch, respectively.

Compared to language-agnostic pruning which degrades performance on certain languages due to a single sub-network, we expect the sparse sub-networks that are tailored to specific languages can mitigate the performance loss. In addition, since the masks can overlap with each other, the ASR pathways share some common parameters between languages, which potentially benefit similar languages (e.g. with shared vocabularies), especially for the low-resource ones.
\begin{table*}[t]
\centering
\resizebox{0.9\textwidth}{!}{
\begin{tabular}{c|c|c|ccccc}
\hline
\multirow{2}{*}{\textbf{Model}} &
  \multirow{2}{*}{\begin{tabular}[c]{@{}c@{}}\textcolor{blue}{\textbf{Monolingual}} or\\ \textcolor{red}{\textbf{Multilingual}}?\end{tabular}} &
  \multirow{2}{*}{\textbf{Sparsity} (\%)} &
  \multicolumn{5}{c}{\textbf{WER} (\%)} \\ \cline{4-8} 
                           &              &      & \textbf{en}             & \textbf{fr}             & \textbf{it}             & \textbf{nl}          & \textbf{Avg.}    \\ \hline \hline
Dense (100M)               & \textcolor{red}{Multilingual} & 0    & 14.03          & 11.74          & 18.90          & 17.69       & 15.59   \\
Dense (30M)                & \textcolor{blue}{Monolingual}  & 0    & 16.13          & 14.64          & 24.66          & 21.64       & 19.27  \\ \hline
LAP                        & \textcolor{red}{Multilingual} & 70.6 & 14.73          & 13.30          & 25.66          & 21.67       & 18.84  \\
LSP - IMP                  & \textcolor{blue}{Monolingual}  & 70.6 & \textbf{11.34} & 11.88          & 18.85          & 18.34       & 15.10  \\
LSP - LTH                  & \textcolor{blue}{Monolingual}  & 70.6 & 11.80          & \textbf{11.85} & \textbf{18.38} & \textbf{18.16} & \textbf{15.05}\\ \hline
ASR Pathways - Random mask & \textcolor{red}{Multilingual} & 70.6 & 13.76          & 13.07          & 20.60          & 19.78          & 16.80\\
ASR Pathways - IMP mask    & \textcolor{red}{Multilingual} & 70.6 & 12.94          & 11.98          & 19.91          & 17.65       &   15.62\\
ASR Pathways - LTH mask &
  \textcolor{red}{Multilingual} &
  70.6 &
  \textbf{12.74} &
  \textbf{11.59} &
  \textbf{17.79} &
  \textbf{17.12} & 
  \textbf{14.81}\\ \hline
\end{tabular}
}
\caption{Test set WERs on MLS. All sparse models have 70.6\% sparsity. The last column shows the average WER on the 4 languages.}
\label{tab:main}
\end{table*}

\vspace{-3mm}
\subsection{Regularization: group lasso weight decay} \label{sec:lasso}
We adopt structured $8\times1$ block-wise pruning. However, the trained dense weights may not follow this pattern. As a result, block-wise pruning may unexpectedly discard important weights, causing performance drop. Hence, it is necessary to have specific algorithmic designs for structured pruning \cite{chen2022coarsening}. Following \cite{liu2022learning}, we use group lasso regularization to mitigate the undesired pruning. Specifically, we define a group as a $8\times1$ block, and the regularization term is added to the loss function $\mathcal{L}$:
\begin{equation} 
\min _W \mathcal{L}+\sum_{i=1}^l \lambda_i \sum_{g \in \mathcal{G}}\left\|W_g^{(i)}\right\|_2 
\end{equation}
where $l$ denotes the number of prunable layers, $W_g^{(i)}$ denotes a weight group in the $i$-th layer, $\lambda_i$ denotes regularization strength. With group lasso, L2-norms of the $8\times1$ blocks will be suppressed, so the blocks become more pruning-friendly: pruning such blocks leads to minimal performance drop. Following \cite{liu2022learning}, we dynamically set $\lambda_i$ based on layer-wise averages of the group L2-norms. We add group lasso regularization in the dense model training stage, as well as in the pruning iterations. Once the model reaches the target sparsity, the regularization is disabled. Since no pruning is involved during or after the multilingual ASR pathways training (Section \ref{sec:pathway}), we do not include the regularization term in this stage. We discuss the effect of group lasso in Section \ref{res:lasso}.

\section{Experimental Setup}
\subsection{Dataset}
We use 4 languages, English (\textit{en}), French (\textit{fr}), Italian (\textit{it}) and Dutch (\textit{nl}) from MLS dataset \cite{pratap2020mls} as our training data.  We follow the train/validation/test splits in \cite{pratap2020mls}. The sizes of the training audio for the 4 languages are 44.7k hrs, 1.1k hrs, 0.2k hrs, 1.6k hrs, respectively. We use 80-dims log-Mel with 25 ms window size and 10 ms step size as the input features. For all multilingual training, we use the sampling strategy in \cite{babu2021xls} with $\alpha=0.5$ to re-balance the training data.

\subsection{Implementation details}
For our dense multilingual model, we use a streaming RNN-T model with 30 Emformer layers with 512 input dims, 2048 feed-forward dims, GeLU non-linearity, and no memory banks \cite{shi2021emformer}, resulting in about 100M parameters. We use an input feature stride 6, Emformer center segment length=4 (240 ms) and right context length=1 (60 ms), which results in 300 ms streaming latency. For the target vocabulary, we generate 512 word pieces for each language, aggregate the word pieces across 4 languages, and remove the duplicated tokens, leading to 1548 items in total. For simplicity, we use the same output layer size for all the multilingual and monolingual training. We use an Adam optimizer with a tri-stage learning rate schedule \cite{park19e_interspeech} and a peak learning rate 1e-3. For monolingual pruning, we set 350K maximum updates and separately tune the pruning interval $T$ (Section \ref{sec:imp_alg}) on the validation set of each language. We train the multilingual ASR pathways for 200K steps. 16 GPUs are used for monolingual training and 32 GPUs for multilingual training, with a per-GPU batch size of 64 samples for both.

\section{Results}

We denote the multilingual language-agnostic pruning as \textit{LAP}, and the monolingual language-specific pruning via IMP and LTH as \textit{LSP - IMP} and \textit{LSP - LTH}, respectively. The proposed ASR pathways trained with IMP masks or LTH masks are denoted as \textit{ASR Pathways - IMP mask} and \textit{ASR Pathways - LTH mask}, respectively. We also include an ASR pathways model trained with randomly initialized language-specific masks (denoted as \textit{ASR Pathways - Random mask}), which serves as a baseline to compare different language-specific masks. All sparse models have 70.6\% sparsity. We denote the pre-trained dense model as \textit{Dense (100M)}. To understand the difference between a sparse model and a small dense model that has around the same number of parameters, we also include a monolingual small dense model baseline by reducing the Emformer layers, which is denoted as \textit{Dense (30M)}. We show the test WERs on the 4 languages in Table \ref{tab:main}.\footnote{Note that the results presented in this paper are all based on low-latency streaming ASR models, so the WERs fall behind \cite{pratap2020mls}, while our in-house non-streaming Transformer/Conformer models \cite{gulati2020conformer, liu2021improving} provide similar WERs compared to \cite{pratap2020mls}.}

\subsection{Language-agnostic vs. language-specific pruning}
First, comparing \textit{Dense (100M)} with \textit{LAP}, we see more significant performance drops on \textit{it} and \textit{nl} than \textit{en} and \textit{fr}, indicating that a single language-agnostic mask does not fit all languages. Since \textit{en} has the most training data, we hypothesize that the learned pruning mask may be biased to \textit{en} and results in performance degradation on other languages. 

Second, we find that both \textit{LSP - IMP} and \textit{LSP - LTH} can perform on par with or better than \textit{Dense (100M)}, even at the high sparsity of over 70\%, suggesting that finding language-specific structures and parameters from an over-parameterized multilingual model with minimal performance drop is possible. Besides, both language-specific pruning methods significantly outperform \textit{Dense (30M)}, which indicates that it is better to prune a pre-trained over-parameterized model than to train a small dense model from scratch.

Third, we compare the proposed ASR pathways trained with different language-specific masks. Both \textit{ASR Pathways - IMP mask} and \textit{ASR Pathways - LTH mask} outperform \textit{ASR Pathways - Random mask}, which suggests that IMP and LTH find better language-specific masks. Note that even with random masks, \textit{ASR Pathways} outperforms \textit{LAP}. This again confirms the importance of having language-specific masks in multilingual ASR scenario. In addition, \textit{ASR Pathways - LTH mask} achieves comparable or better performance compared to the monolingual \textit{LSP - LTH} on most languages (except for \textit{en}), meaning that our proposed multilingual sparse ASR pathways can match the performance of monolingual sparse models. The improvements on \textit{fr}, \textit{it} and \textit{nl} indicate the benefit of multilingual joint training on low-resource languages. On the flip side, the performance gap on \textit{en} might be an outcome of the language interference problem \cite{lu2022language}: high-resource languages can often suffer from negative transfer during multilingual joint training. The high sparsity makes it an even more challenging learning task. Finally, we observe that using LTH masks lead to better performance than IMP masks, consistent with the finding in \cite{ding2021audio}.

\begin{figure}[t]
    \centering
    \includegraphics[width=1.0\columnwidth]{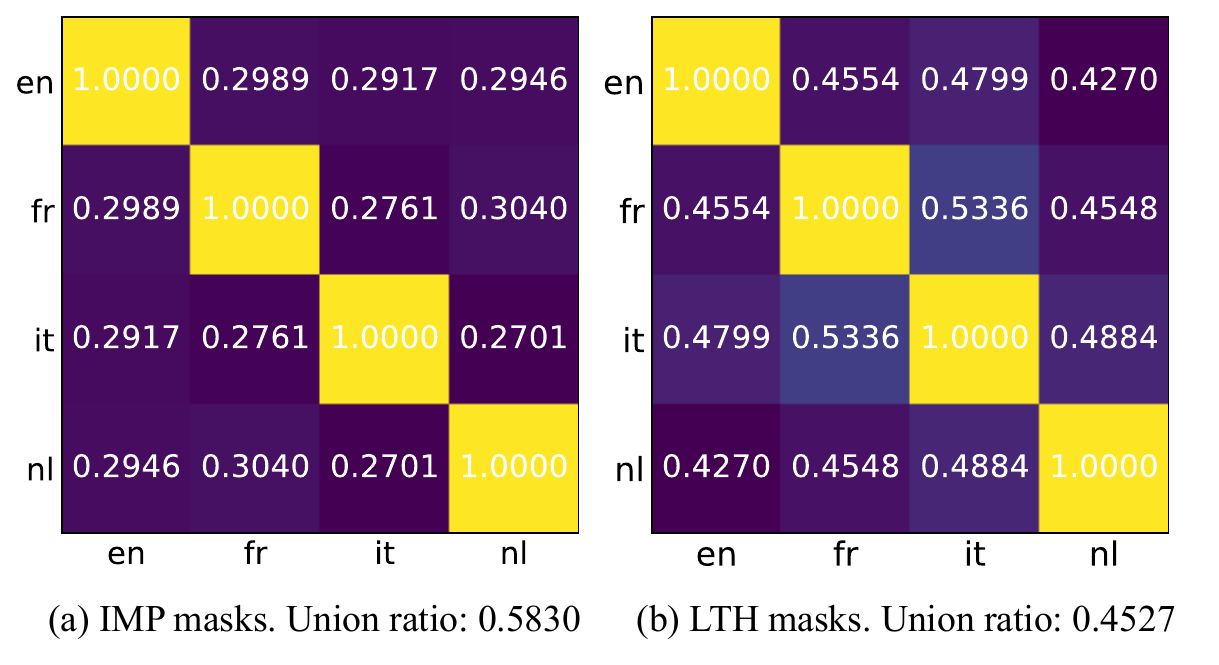}
    \vspace{-7mm}
    \caption{IOUs and union ratios of (a) IMP masks and (b) LTH masks.}
    \label{fig:masks}
\end{figure}

\subsection{Analysis on the language-specific masks} \label{analyze_mask}
We conduct analyses on the learned language-specific masks to understand their distributions across different languages. We compute Intersection Over Union (IOU) between a mask pair $(m_i, m_j)$:
\begin{equation}
    IOU(m_i, m_j) = \frac{|m_i=1 \cap m_j=1|}{|m_i=1 \cup m_j=1|}
\end{equation}
A higher IOU suggests more overlapping between the surviving parameters of two language-specific masks. In addition, we define union ratio (UR) of the four masks $(m_1, ..., m_4)$ as
\begin{equation}
    UR(m_1, ..., m_4) = \frac{|\bigcup_{n=1}^4 m_n=1|}{|m_1|}
\end{equation}
which measures how the union of the surviving parameters of the sub-networks span the whole parameters space. A higher UR suggests a higher total effective parameters usage.

We show the IOUs and union ratios in Figure \ref{fig:masks}. Comparing LTH masks with IMP masks, we find significantly higher IOUs in LTH masks. This indicates that LTH masks have more overlaps between language pairs, resulting in more shared parameters across different languages. As a result, the union ratio is also lower than IMP masks, meaning that LTH masks are using less total effective model parameters. One possible reason is that, in each iteration LTH starts from the same model parameters ($\theta_0$ in Section \ref{sec:imp_alg}), while IMP starts from different parameters ($\theta_T$ in Section \ref{sec:imp_alg}), making LTH better at keeping the universal part of parameters in $\theta_0$ that are useful for all languages. In contrast, IMP will gradually lose this information and learn more language-specific parameters, producing less mask overlaps. Note that even with a lower union ratio (smaller total parameters usage), LTH masks can outperform IMP masks.

With a union ratio at 0.4527, \textit{ASR Pathways - LTH mask} is essentially having more effective parameters for the 4 languages compared to the 70.6\%-sparse \textit{LAP}. In order to have a fairer comparison to \textit{LAP}, we also train a \textit{LAP} model with 56.5\% sparsity and obtain test WERs of 13.93 (\textit{en}), 12.58 (\textit{fr}), 26.57 (\textit{it}), 26.09 (\textit{nl}). Corroborating with the results in Table \ref{tab:main}, the overall trend still holds, confirming the superiority of language-specific masks.

\subsection{Effect of group lasso regularization} \label{res:lasso}

\begin{table}[t]
\resizebox{1.0\columnwidth}{!}{
\begin{tabular}{l|cccc}
\hline
\multicolumn{1}{c|}{\multirow{2}{*}{\textbf{Model}}} & \multicolumn{4}{c}{\textbf{WER} (\%)}                                      \\ \cline{2-5} 
\multicolumn{1}{c|}{}                       & \textbf{en}             & \textbf{fr}             & \textbf{it}             & \textbf{nl}             \\ \hline
Dense (100M)                               & \textbf{13.48} & \textbf{11.27} & 21.14          & 19.55          \\
\quad\quad + group lasso                              & 14.03          & 11.74          & \textbf{18.90} & \textbf{17.69} \\ \hline
LSP - IMP                                  & 13.01          & 13.28          & 20.94          & 20.93          \\
\quad\quad + group lasso                              & \textbf{11.34} & \textbf{11.88} & \textbf{18.85} & \textbf{18.34} \\
LSP - LTH                                  & 14.80          & 18.50          & 28.87          & 24.92          \\
\quad\quad + group lasso                              & \textbf{11.80} & \textbf{11.85} & \textbf{18.38} & \textbf{18.16} \\ \hline
ASR Pathways - IMP mask                         & 14.05          & 13.20          & 20.15          & 18.87          \\
\quad\quad + group lasso                              & \textbf{12.94} & \textbf{11.98} & \textbf{19.91} & \textbf{17.65} \\
ASR Pathways - LTH mask                         & 14.65          & 14.04          & 21.67          & 20.77          \\
\quad\quad + group lasso                              & \textbf{12.74} & \textbf{11.59} & \textbf{17.79} & \textbf{17.12} \\ \hline
\end{tabular}
}

\caption{Test set WERs with and without group lasso (Section \ref{sec:lasso}).}
\label{tab:lasso}
\end{table}
In this section, we discuss the effect of the group lasso regularization (Section \ref{sec:lasso}). We pre-train two dense models (i.e. $\theta_0$ in Section \ref{sec:imp_alg}) with and without group lasso regularization. Then we compare the performance of the sparse models using the two dense models as the pruning starting point, respectively.\footnote{Note that for best pruning performance we enable the regularization for both during pruning.} Table \ref{tab:lasso} shows test set WERs, where the ``+ group lasso" rows denote models based on the regularization-enabled dense model (same as in Table \ref{tab:main}). With group lasso, \textit{LSP - IMP} and \textit{LSP - LTH} gain significant improvements. Correspondingly, the ASR pathways also show significant boosts. Interestingly, without group lasso, the performance of LTH largely falls behind IMP, and group lasso appears to have a larger improvement for LTH than IMP. As we analyzed in Section \ref{analyze_mask}, due to the rewinding step, the initialization $\theta_0$ has a major impact on LTH. Hence, it is essential to have structured-pruning-friendly pre-trained dense weights. 

\section{Conclusion}

We have presented multilingual ASR pathways - a design that comes with language-specific activation. We show that such a sparse multilingual ASR model outperforms both dense model and language-agnostic pruning, and has better performance on low-resource languages compared to the monolingual sparse models. Further, via our analyses on the language-specific masks, we reveal that with the help of the group lasso regularization, LTH masks are superior to IMP masks, even with less total effective parameters. In the future, we plan to extend ASR pathways to more languages and different tasks.

\vfill\pagebreak

\bibliographystyle{IEEEbib}
\bibliography{refs}

\begin{thebibliography}{10}

\bibitem{movsner2019improving}
Ladislav Mo{\v{s}}ner, Minhua Wu, Anirudh Raju, Sree Hari~Krishnan
  Parthasarathi, Kenichi Kumatani, et~al.,
\newblock ``Improving noise robustness of automatic speech recognition via
  parallel data and teacher-student learning,''
\newblock in {\em Proc. ICASSP}, 2019.

\bibitem{panchapagesan2021efficient}
Sankaran Panchapagesan, Daniel~S Park, Chung-Cheng Chiu, et~al.,
\newblock ``Efficient knowledge distillation for rnn-transducer models,''
\newblock in {\em Proc. ICASSP}, 2021.

\bibitem{shangguan2019optimizing}
Yuan Shangguan, Jian Li, Qiao Liang, et~al.,
\newblock ``Optimizing speech recognition for the edge,''
\newblock {\em arXiv preprint arXiv:1909.12408}, 2019.

\bibitem{yang2022omni}
Haichuan Yang, Yuan Shangguan, Dilin Wang, et~al.,
\newblock ``Omni-sparsity dnn: Fast sparsity optimization for on-device
  streaming e2e asr via supernet,''
\newblock in {\em Proc. ICASSP}, 2022.

\bibitem{ding2021audio}
Shaojin Ding, Tianlong Chen, and Zhangyang Wang,
\newblock ``Audio lottery: Speech recognition made ultra-lightweight,
  noise-robust, and transferable,''
\newblock in {\em Proc. ICLR}, 2022.

\bibitem{liu2022learning}
Chunxi Liu, Yuan Shangguan, Haichuan Yang, Yangyang Shi, Raghuraman
  Krishnamoorthi, and Ozlem Kalinli,
\newblock ``Learning a dual-mode speech recognition model via self-pruning,''
\newblock in {\em Proc. SLT}, 2022.

\bibitem{he2019streaming}
Yanzhang He, Tara~N. Sainath, Rohit Prabhavalkar, et~al.,
\newblock ``Streaming end-to-end speech recognition for mobile devices,''
\newblock in {\em Proc. ICASSP}, 2019.

\bibitem{sainath2020streaming}
Tara~N. Sainath, Yanzhang He, Bo~Li, et~al.,
\newblock ``A streaming on-device end-to-end model surpassing server-side
  conventional model quality and latency,''
\newblock in {\em Proc. ICASSP}, 2020.

\bibitem{ding20224}
Shaojin Ding, Phoenix Meadowlark, Yanzhang He, et~al.,
\newblock ``4-bit conformer with native quantization aware training for speech
  recognition,''
\newblock {\em arXiv preprint arXiv:2203.15952}, 2022.

\bibitem{toshniwal2018multilingual}
Shubham Toshniwal, Tara~N. Sainath, Ron~J. Weiss, et~al.,
\newblock ``Multilingual speech recognition with a single end-to-end model,''
\newblock in {\em Proc. ICASSP}, 2018.

\bibitem{pratap20c_interspeech}
Vineel Pratap, Anuroop Sriram, Paden Tomasello, et~al.,
\newblock ``{Massively Multilingual ASR: 50 Languages, 1 Model, 1 Billion
  Parameters},''
\newblock in {\em Proc. Interspeech}, 2020.

\bibitem{li2021scaling}
Bo~Li, Ruoming Pang, Tara~N. Sainath, et~al.,
\newblock ``Scaling end-to-end models for large-scale multilingual asr,''
\newblock in {\em Proc. ASRU}, 2021.

\bibitem{kannan19_interspeech}
Anjuli Kannan, Arindrima Datta, Tara~N. Sainath, et~al.,
\newblock ``{Large-Scale Multilingual Speech Recognition with a Streaming
  End-to-End Model},''
\newblock in {\em Proc. Interspeech}, 2019.

\bibitem{bai2022joint}
Junwen Bai, Bo~Li, Yu~Zhang, et~al.,
\newblock ``Joint unsupervised and supervised training for multilingual asr,''
\newblock in {\em Proc. ICASSP}, 2022.

\bibitem{lu2022language}
Yizhou Lu, Mingkun Huang, Xinghua Qu, et~al.,
\newblock ``Language adaptive cross-lingual speech representation learning with
  sparse sharing sub-networks,''
\newblock in {\em Proc. ICASSP}, 2022.

\bibitem{gaur2021mixture}
Neeraj Gaur, Brian Farris, Parisa Haghani, et~al.,
\newblock ``Mixture of informed experts for multilingual speech recognition,''
\newblock in {\em Proc. ICASSP}, 2021.

\bibitem{dean_2021}
Jeff Dean,
\newblock ``Introducing pathways: A next-generation ai architecture,''
  \url{https://blog.google/technology/ai/introducing-pathways-next-generation-ai-architecture/},
\newblock accessed: Sep-2022.

\bibitem{zhu2017prune}
Michael Zhu and Suyog Gupta,
\newblock ``To prune, or not to prune: exploring the efficacy of pruning for
  model compression,''
\newblock {\em arXiv preprint arXiv:1710.01878}, 2017.

\bibitem{frankle2018lottery}
Jonathan Frankle and Michael Carbin,
\newblock ``The lottery ticket hypothesis: Finding sparse, trainable neural
  networks,''
\newblock in {\em Proc. ICLR}, 2019.

\bibitem{shi2021emformer}
Yangyang Shi, Yongqiang Wang, Chunyang Wu, Ching-Feng Yeh, Julian Chan, et~al.,
\newblock ``Emformer: Efficient memory transformer based acoustic model for low
  latency streaming speech recognition,''
\newblock in {\em Proc. ICASSP}, 2021.

\bibitem{shi2022streaming}
Yangyang Shi, Chunyang Wu, Dilin Wang, Alex Xiao, et~al.,
\newblock ``Streaming transformer transducer based speech recognition using
  non-causal convolution,''
\newblock in {\em Proc. ICASSP}, 2022.

\bibitem{graves2012sequence}
Alex Graves,
\newblock ``Sequence transduction with recurrent neural networks,''
\newblock {\em arXiv preprint arXiv:1211.3711}, 2012.

\bibitem{narang2017block}
Sharan Narang, Eric Undersander, and Gregory Diamos,
\newblock ``Block-sparse recurrent neural networks,''
\newblock {\em arXiv preprint arXiv:1711.02782}, 2017.

\bibitem{tian2022lae}
Jinchuan Tian, Jianwei Yu, Chunlei Zhang, Chao Weng, Yuexian Zou, and Dong Yu,
\newblock ``Lae: Language-aware encoder for monolingual and multilingual asr,''
\newblock {\em arXiv preprint arXiv:2206.02093}, 2022.

\bibitem{zhu20d_interspeech}
Yun Zhu, Parisa Haghani, Anshuman Tripathi, et~al.,
\newblock ``{Multilingual Speech Recognition with Self-Attention Structured
  Parameterization},''
\newblock in {\em Proc. Interspeech}, 2020.

\bibitem{conneau2020unsupervised}
Alexis Conneau, Alexei Baevski, Ronan Collobert, Abdelrahman Mohamed, and
  Michael Auli,
\newblock ``Unsupervised cross-lingual representation learning for speech
  recognition,''
\newblock {\em arXiv preprint arXiv:2006.13979}, 2020.

\bibitem{wu2021dynamic}
Zhaofeng Wu, Ding Zhao, Qiao Liang, Jiahui Yu, Anmol Gulati, and Ruoming Pang,
\newblock ``Dynamic sparsity neural networks for automatic speech
  recognition,''
\newblock in {\em Proc. ICASSP}, 2021.

\bibitem{sun2020learning}
Tianxiang Sun, Yunfan Shao, Xiaonan Li, Pengfei Liu, Hang Yan, Xipeng Qiu, and
  Xuanjing Huang,
\newblock ``Learning sparse sharing architectures for multiple tasks,''
\newblock in {\em Proc. AAAI}, 2020.

\bibitem{chen2020lottery}
Tianlong Chen, Jonathan Frankle, Shiyu Chang, Sijia Liu, Yang Zhang, et~al.,
\newblock ``The lottery ticket hypothesis for pre-trained bert networks,''
\newblock in {\em Proc. NeurIPS}, 2020.

\bibitem{chen2022coarsening}
Tianlong Chen, Xuxi Chen, Xiaolong Ma, Yanzhi Wang, and Zhangyang Wang,
\newblock ``Coarsening the granularity: Towards structurally sparse lottery
  tickets,''
\newblock {\em arXiv preprint arXiv:2202.04736}, 2022.

\bibitem{pratap2020mls}
Vineel Pratap, Qiantong Xu, Anuroop Sriram, Gabriel Synnaeve, et~al.,
\newblock ``{MLS}: A large-scale multilingual dataset for speech research,''
\newblock {\em arXiv preprint arXiv:2012.03411}, 2020.

\bibitem{babu2021xls}
Arun Babu, Changhan Wang, Andros Tjandra, Kushal Lakhotia, Qiantong Xu, et~al.,
\newblock ``Xls-r: Self-supervised cross-lingual speech representation learning
  at scale,''
\newblock {\em arXiv preprint arXiv:2111.09296}, 2021.

\bibitem{park19e_interspeech}
Daniel~S. Park, William Chan, Yu~Zhang, et~al.,
\newblock ``{SpecAugment: A Simple Data Augmentation Method for Automatic
  Speech Recognition},''
\newblock in {\em Proc. Interspeech}, 2019.

\bibitem{gulati2020conformer}
Anmol Gulati, James Qin, Chung-Cheng Chiu, Niki Parmar, Yu~Zhang, et~al.,
\newblock ``Conformer: Convolution-augmented transformer for speech
  recognition,''
\newblock {\em Proc. Interspeech}, 2020.

\bibitem{liu2021improving}
Chunxi Liu, Frank Zhang, Duc Le, Suyoun Kim, Yatharth Saraf, and Geoffrey
  Zweig,
\newblock ``Improving rnn transducer based asr with auxiliary tasks,''
\newblock in {\em Proc. SLT}, 2021.

\end{thebibliography}

\end{document}